\def \abc#1#2#3#4 {\reference#1, {\sl#2}, {\bf#3}, #4}
\def \blank {\lower 5pt\hbox to 0.75in{\hrulefill}}

\def \cm{~\rm{cm}}
\def \s{~\rm{s}}
\def \km{~\rm{km}}

\def \K{~\rm{K}}

\def \yr{~\rm{yr}}

\def \pc{~\rm{pc}}
\def \kpc{~\rm{kpc}}
\documentclass[12pt,preprint]{aastex}
\begin{document}

\title{THE EARLY INTERACTION OF THE PLANETARY NEBULA NGC 40 WITH THE INTERSTELLAR MEDIUM}

\author{Jodie Martin, Kiriaki Xilouris, and  Noam Soker\altaffilmark{1}} 

\affil{University of Virginia, Department of Astronomy, P.O.~Box 3818, 
Charlottesville, VA 22903-0818; 
jrm7w@virginia.edu; kx8u@virginia.edu; soker@physics.technion.ac.il }

\altaffiltext{1}{ on sabbatical from the Department of Physics,
University of Haifa at Oranim, Israel} 

\begin{abstract}

We report the detection of an H$\alpha$ emission-line structure in the 
upstream side of the planetary nebula NGC 40, which is predicted by numerical 
simulations, and which is attributed to Rayleigh-Taylor instability.
Such a  Rayleigh-Taylor instability is expected to occur at early 
stages of the interaction process between the interstellar medium (ISM)
and a fast moving planetary nebula, as is the case for NGC 40.
We resolved the Rayleigh-Taylor instability `tongues', as well
as the flatness of the nebula around the  `tongues', which results 
from the deceleration by the ISM.
 
\end{abstract}


\section{INTRODUCTION}

Planetary nebulae (PNe) are expanding ionized nebulae expelled
from asymptotic giant branch (AGB) stars. 
The remnant of the AGB star, i.e., its core, ionizes the nebula as it
evolves to become a white dwarf. 
As the gas expands, its density drops and its expansion starts to
be influenced by the ISM.
If there is a relative motion between the PN and the ISM,
the ISM signature on the PN structure will emerge first on the nebular side
facing the ISM, i.e., the upstream direction, 
where the ISM ram pressure on the nebula is the largest.
At late times, when the nebula is relatively large and of low density,
this type of interaction forms a bow-shaped nebula, e.g., Sh 2-216 
(Tweedy, Martos,\& Noriega-Crespo 1995), KeWe 5 (Kerber et al.\ 1998),
KFR 1 (Rauch et al.\ 2000), and SB 51 (PN G 357.4-07.2;
Beaulieu, Dopita, \& Freeman 1999),
sometimes with a long extension of the bow-shaped region on the
down stream direction, e.g. Abell 35 (Jacoby 1981; Hollis et al.\ 1996).
Since the nebula is large and of low surface brightness at this late 
stage, telescopes with large fields of view should be used to detect
these PNe, as was done by Xilouris et al.\ (1996) and Tweedy \& Kwitter 
(1996) in their study of many ancient PNe interacting with the ISM. 
Due to these and other (e.g., Borkowski, Tsvetanov, \& Harrington 1993;
Tweedy \& Kwitter 1994;  Muthu, Anandarao, \& Pottasch 2000; 
Kerber et al.\ 2000, 2002) 
observational studies, and several theoretical works 
(e.g., Borkowski, Sarazin, \& Soker 1990; 
Soker, Borkowski, \& Sarazin 1991, hereafter SBS;
Villaver, Manchado, \& Garc\'ia-Segura 2000; Dopita et al.\ 2000) 
the basic physics of this stage of PN-ISM interaction is well understood 
and there are many examples of PNe evolving through this stage. 
 
The PN-ISM interaction process may be influenced 
by the ISM magnetic field (e.g., Soker \& Dgani 1997; Soker \& Zucker 1997) 
and several types of instabilities (e.g., SBS;  
Dgani \& Soker 1994; see review by Dgani 2000). 
Of particular interest is the Rayleigh-Taylor (RT) instability, which may 
fragment the outer halo, hence allowing the ISM to stream and interact 
with nebular material closer to the central star (Dgani \& Soker 1998). 
At early stages the RT instability manifests itself as
`tongues' located outside, but connected to, the main nebular shell
(e.g., SBS; Dopita et al.\ 2000).
This is a general type of structure found in simulations of other dense clouds
moving through a low density medium (e.g., Jones, Kang, \& Tregillis 1994). 
To our best knowledge, no such RT `tongues' were reported
before for PNe interacting with the ISM. 
The only claim for a RT instability at an early stage of PN-ISM interaction
was made by Zucker \& Soker (1993; hereafter ZS93) for IC 4593. 
IN IC 4593 the nebula is compressed in the upstream direction, such that its
density becomes higher, and there is a structure protruding from the nebula
in the upstream direction.
ZS93 term the region which protrudes from the nebula a `bump'.
However, in IC 4593 the `bump' is blobby, and does not show any 
large scale internal structure, i.e., no RT tongues are observed.

In the present paper we report the detection of a structure in the upstream 
side of the interacting PN NGC 40 (PN G120.0+09.8), 
which closely resembles the structure of RT instability obtained in 
numerical simulations. 
 We focus on the RT instability feature, which following ZS93 we term
a `bump', although its internal structure reveals features not seen
in IC 4593, and which are predicted by numerical simulations. 
It is quite possible that in NGC 40 the interaction process is at an
earlier stage than that of IC 4593. 
 Other properties of NGC 40 are thoroughly studied by Meaburn
et al.\ (1996), and hence will not be discussed here. 
In addition to strengthening the theory of PN-ISM interaction, we
hope that this first example of a RT instability at very early stages of
PN-ISM interaction will stimulate further observations in searching
for such features.    

\section{OBSERVATIONS}
  
Thirty minute exposures were taken of NGC 40 in two narrow band 
interference filters with the Fan Mountain $40\arcsec$ Astrometric 
Telescope on the night of December 4, 2001.  
The two filters are centered at $\lambda 5020~$\AA,
to isolate [OIII]$~\lambda 5007~$\AA ~emission ([OIII] filter),
and $\lambda 6575~$\AA, to isolate H$\alpha~\lambda~6563~$\AA 
and [NII]$~\lambda 6584~$\AA~emission (H$\alpha$ filter). 
The FWHMs (full width at half maximum) of the filters are 
$53~$\AA ~and $68~$\AA, respectively, 
and both filters have a throughput of about $70\%$. 
The detector is a San Diego State/SITE $2048\times2048$ Engineering 
grade CCD, and each pixel has a spatial resolution of $0\farcs36$.  
The images were reduced using the standard procedures of the IRAF data 
reduction software.  Bias subtraction, dome flat field correction, trimsec, 
and biassec were done.
As stated above, we focus only on the outer regions which are dominated by 
the H$\alpha$ and [NII] emission, so only the H$\alpha$ filter image results 
will be discussed.
The effective seeing of the H$\alpha$ filter image is $1\farcs8$ FWHM and 
the mean airmass during the exposure was 1.21.  
The IDL routine, `sky', was used to determine the value of the sky in 
units of {$ADU/pixel$} on a region of the chip that is dominated by 
background and is free of vignetting effects.  
The image was normalized by subtracting the value of the sky.  
The subtracted image was then median smoothed by 8 pixels with even 
interpolation.  The median smoothed, subtracted image has a sky value of 
{$0\pm1ADU/pixel$}.
Our results are presented in Figures 1a-1e, where the intensity scale 
is linear.  
 
The RT instability feature is seen on the northwest side of the main 
(bright) nebular shell.  
Following ZS93 we term this region the `bump'.  
The ram pressure of the ISM decelerates the expanding gas facing 
the ISM and flattens the nebula as can be seen in the extended 
isophotes of Figures 1b-1e.  
The limb brightened regions of the bump which are attached to the 
main nebula, and which are termed `RT-tongues', are indicated in 
Figure 1b, as well as the bright clumps that are observed
at the front of the bump. 
Note that the other `blobs' and `filaments' in the images which
are scattered around the main nebula have coherence scales much smaller 
than the size of the `bump' and the flatness to its sides.
Such `blobs' are expected to be scattered around PNe, in particular
PNe interacting with the ISM (ZS93).
The bright blob adjacent to the southwestern edge of the nebula are two
close background and/or foreground stars.
 
Figure 1a shows the light profile along a cut through the bump and
between the RT-tongues, as indicated in Figure 1b.  
The cut distinctly shows a minimum at {$r=56\farcs5$},  
a sharp rise again when it encounters the clumps, 
and a return to the background as the cut moves into the surrounding ISM.  
Hints of the RT-tongues are seen in Figures  2a and 2c of 
Meaburn et al.\ (1996).  
The bent northern jet, a condensation,  which may result from the 
interaction with the ISM, and the direction of proper motion are 
indicated in Figure 1d. 
Note that the condensation (`Cond' in Fig. 1d) does not have the point spread 
function of the known stars in the field. 
The proper motion of the central star was measured by the HIPPARCOS 
mission (Perryman et al.\ 1997) to be 
${\mu_{\alpha}}=-8.29\pm4.92~$mas~yr$^{-1}$, and 
${\mu_{\delta}}=4.36\pm4.36~$mas~yr$^{-1}$. 
The thick arrow indicates the proper motion, while the two thin arrows 
indicate the extreme {$1\sigma$} limits of the total proper motion direction.  
 
To demonstrate the similarity of the RT-bump, RT-tongues, and flatness, 
to the structure obtained in 2D numerical simulations, we show in Figure 1f 
the results from SBS, where more details are given 
(see also Jones et al.\ 1994 for a very similar RT instability structure).  
Note that the numerical results present the density in the symmetry plane, 
while the observations are the integrated emission along line of sight 
through the nebula.  
Hence, there is projected emission near the symmetry axis of the 
interaction flow (tip of the bump) in the H$\alpha$ images, although no 
dense material is located there in the simulations.

\section{INTERACTION WITH THE ISM}

 Prior to our work, there were two indications for the interaction 
of NGC 40 with the ISM: (i) The bent northern jet (for its detailed 
structure see Meaburn et al.\ 1996),  and (ii) condensations 
(dense clumps) and filaments around the nebula, which are thought
to be formed from instabilities of the outer low density halo as it
interacts with the ISM (ZS93). 
The proper motion is more or less along the direction expected from
the bending angle of the northern jet. 
These features remind us of IC 4593, which has its northern FLIER
(fast low-ionization emission regions along the symmetry axis) closer
to the central star than the southern FLIER (ZS93; Corradi et al.\ 1996), 
and condensations outside the main nebula (ZS93).  
We were not surprised therefore to find a `bump' in the direction of 
proper motion (more or less), as in IC 4593 (ZS93). 
Whereas in IC 4593 the bump is only suggestive for a RT instability,
here the similarity between the bump and the numerical simulations is obvious.
For comparison we present in Figure 1f the density contours map in the 
symmetry plane from the 2D simulations of SBS.
In that figure the nebula is moving upward, with the left edge of the panel
being the symmetry axis. 
 The RT instability and the flatness are clearly seen; 
a similar structure with higher numerical resolution, can be seen 
in the numerical simulations of a dense cloud moving through a low 
density medium (Jones et al.\ 1994, their fig. 1a). 
We see in NGC 40 (Figs 1b-1e) both the RT tongues (which in projection 
appear as a limb brightened bump) and the flatness of the nebular edge in
the upstream direction; this flatness results from the deceleration
of the nebular material in the upstream side.  
We note that the size of the bump is about equal to the deceleration
distance of the upstream side, as expected at early stages (SBS).
We conclude that the bump is due to RT instability.

The condition for the development of the RT instability is that the ISM 
density be lower than that of the nebula; for that the ISM should be hotter
than the nebula (SBS; we follow their analysis).
SBS assume that the nebula is at a temperature of $\sim 10^4 \K$, and 
calculate the cooling time of the hot shocked ISM to that temperature (SBS).
The cooling time depends on the relative velocity of the ISM and the nebula,
and on the density and ionization stage of the ISM. 
For the angular proper motion given in the previous section and a distance of
$980\pm100 \pc$ to NGC 40 (Bianchi and Grewing, 1987), 
the transverse motion is $v_t \sim 44 \pm 30 \km \s^{-1}$.
The heliocentric radial velocity of NGC 40 is $v_r = - 32 \km \s^{-1}$
(Meaburn et al.\ 1996).
The absolute value of the velocity of NGC 40 relative to the sun is 
therefore $v_\ast = (v_r^2+v_t^2)^{1/2} \simeq 54 \km \s^{-1}$.
 For a nebular expansion velocity of $v_e=10 \km \s^{-1}$, and neglecting the
ISM motion, we find for the relative velocity between the nebular gas 
and the ISM, $v_{\rm rel} = v_\ast+v_e \sim 64 \km \s^{-1}$. 
We can safely take then $v_{\rm rel} = 60 \km \s^{-1}$.
We assume that NGC 40 is moving in the warm ionized medium, for which the 
average hydrogen number density at the location of NGC 40, namely
$170 \pc$ from the galactic plane, is $< 0.05 \cm^{-3}$ (Reynolds 1989).    
The ISM is shocked to a temperature of $\sim 5 \times 10^4 \K$, from which
the cooling time to a temperature of $\sim 10^4 \K$ is 
$t_{\rm cool} \sim 3 \times 10^4 (n_o/0.05 \cm^{-3})^{-1} \yr$. 
 The radius of the NGC 40 (at the location of the bump) is $1^{\prime}$,
or $R_{\rm PN} = 9 \times 10^{17} \cm$, assuming a distance of
$1 \kpc$ to NGC 40. 
The flow time of the ISM along a distance of $R_{\rm PN}$,
and for $v_\ast=50 \km \s^{-1}$, is $t_{\rm flow} \sim 6 \times 10^3 \yr$.
 Since $t_{\rm cool} > t_{\rm flow}$, the ISM has no time to cool
and an adiabatic flow, prone to RT instability, commences (SBS).
 In addition, at a very early stage, before the nebula is ionized, its
temperature is $\lesssim 10^3 \K$, so even if the ISM cools faster down to
$\sim 10^4 \K$, its cooling will be slower below that temperature, and the
ISM will be hotter and less dense than the nebular gas
at early stages.  
 The expansion time of the material in the bump, since it left the central
star, is $ R_{\rm pn}/v_e \simeq  3 \times 10^4 \yr$, long enough for
the development of a RT instability bump (SBS).

 To conclude, NGC 40 is moving through, and interacting with, the ISM.
The motion and indication for this interaction were known previously.
We report for the first time the detection of a structure on the upstream
edge of the main nebula which is predicted by numerical simulations,
and which is attributed to RT instability at an
early stage of the interaction process. 
Such a RT instability is expected to occur at early stages of the 
PN-ISM interaction under conditions similar to those for NGC 40.

\bigskip

{\bf ACKNOWLEDGMENTS:}
We thank Robert O'Connell for valuable comments on the original manuscript. 
J. M. was supported by the NASA Long Term Space Astrophysics
grant NAG5-6403.   
N. S. was supported by a Celerity Foundation Distinguished Visiting 
Scholar grant at the University of Virginia, and by a grant from the
US-Israel Binational Science Foundation.


{\bf Fig1}  Mosaic figure emphasizing the Rayleigh-Taylor (RT) instability  
`bump', and its internal structure with the tongues and clumps.  
North is up and East is to the left.  
These features, which are defined on panel $b$ and $d$, can be clearly   
seen in panels $c$ and $e$.  
($a$)  H$\alpha$ filter intensity vs. distance from the central star in 
units of arcsec, and along the radial cut marked on panel $b$.  
The intensity scale in the images is linear in units of {$ADU/pixel$}.
($b$) H$\alpha$ filter surface brightness contour map of the RT `bump' 
region.    
The contours have the values (same scale as in panel $a$) of 9 (dark line), 
10 (medium line), and 11 (light line).   
Axes are distance from the central star in arcsec. ($c$) Gray scale image 
($180\arcsec\times 180\arcsec$). 
Image is scaled between the values 5 and 50 {$ADU/pixel$}.  
($d$) H$\alpha$ filter surface brightness contour map of $c$.   
The contours have the values of 9, 10, 11, 25, 200, 700, 2000, and 6000.   
The arrows indicate the direction of the proper motion of the central star.   
The bold arrow is the direction measured by HIPPARCOS and the lighter  
arrows indicate the extreme {$1\sigma$} directions given by the errors  
in the HIPPARCOS measurement. 
`Cond' stands for a condensation.   
Units on the axes are arcsec.  
($e$) Gray scale of the RT `bump' region of $b$.  
Again, the image is scaled between the values 5 and 50 {$ADU/pixel$}.
($f$) Density contour map for an adiabatic simulation of a dense PN shell 
from Soker et al.\ (1991).   
The nebula is moving upward in this panel, with the left edge of the panel 
being the symmetry axis of the flow.   
Contours marked 0.006 and 0.01 indicate the location of the ISM shock front.  
Units on the axes are $10^{17}$ cm, and density levels are in units of  
$10^{-24}$ g cm$^{-1}$.


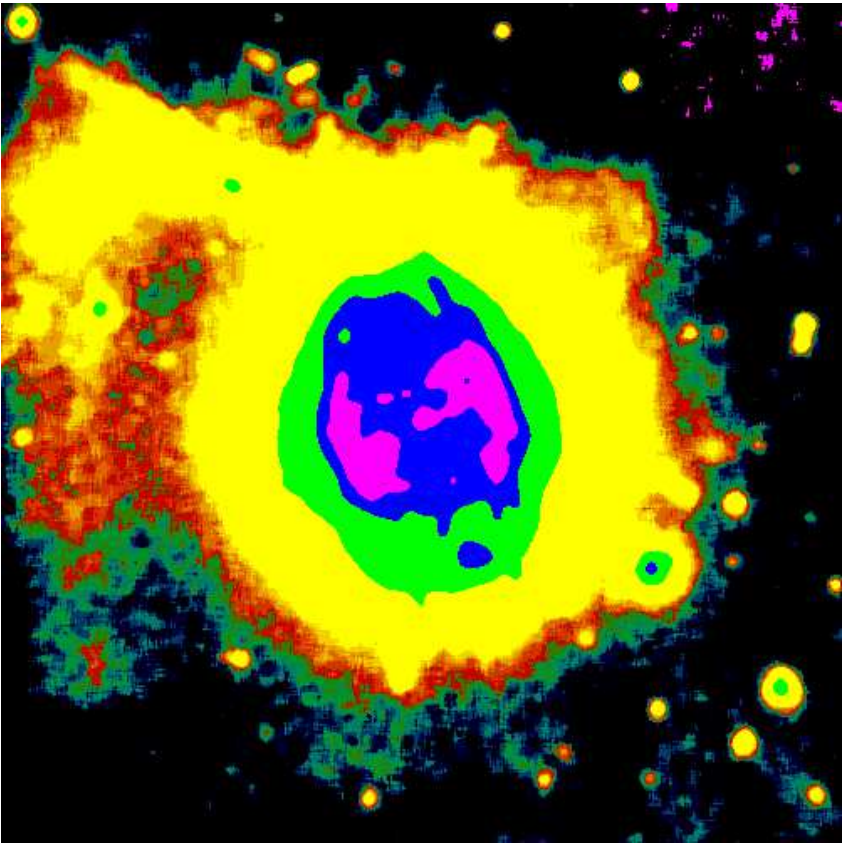
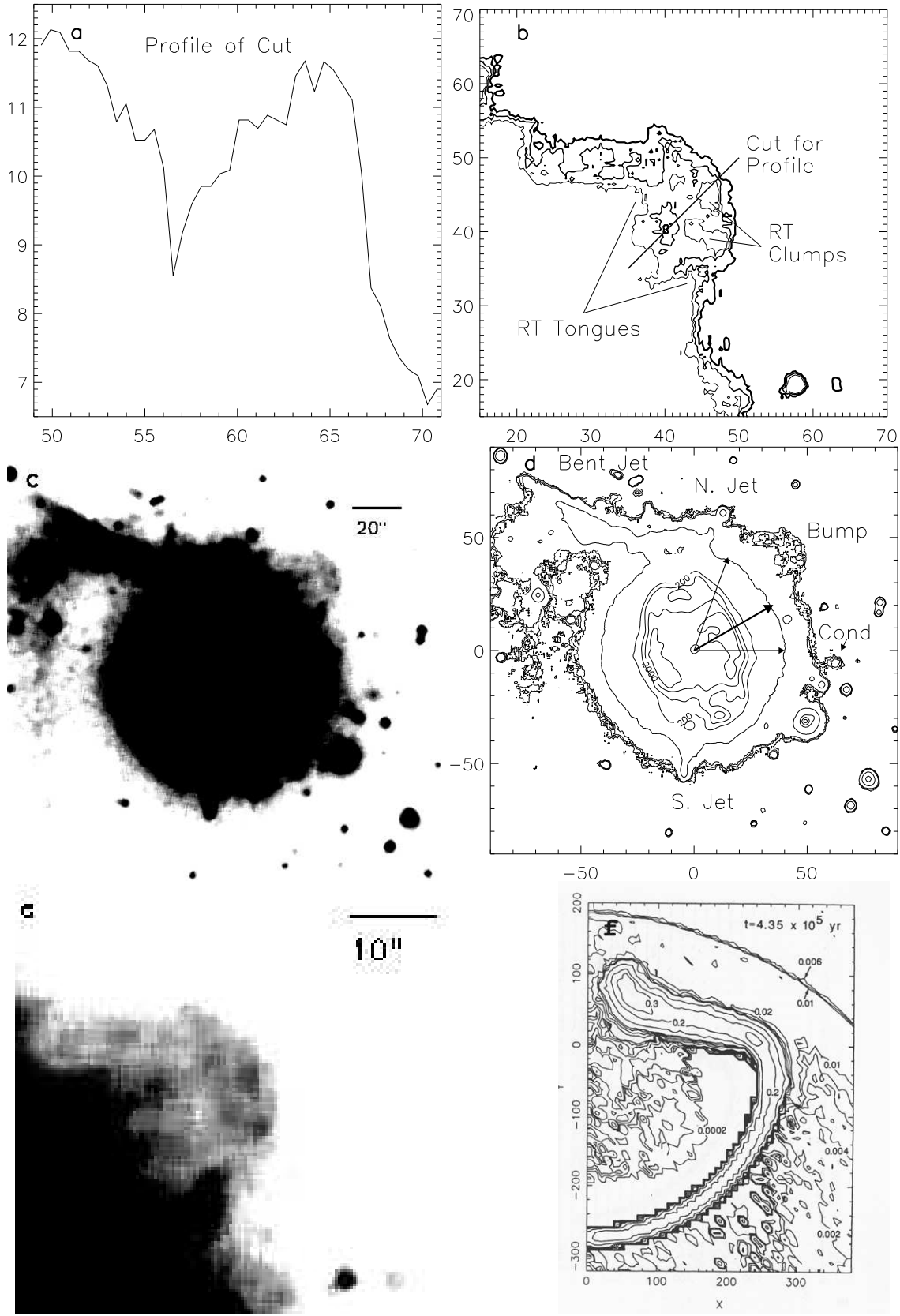
\end{document}